\documentclass[conference]{IEEEtran}

\usepackage{amsmath,amssymb,amsfonts}
\usepackage[algo2e,ruled,lined,boxed,commentsnumbered, noend, linesnumbered, procnumbered]{algorithm2e}
\usepackage[binary-units]{siunitx}
\usepackage[abbreviations]{foreign}
\usepackage{graphicx}
\usepackage{textcomp}
\usepackage[dvipsnames]{xcolor}
\usepackage{xspace}
\usepackage{booktabs} 
\usepackage{multirow}
\usepackage[backend=bibtex,sorting=none,style=ieee,isbn=false]{biblatex}
\usepackage[hidelinks]{hyperref}
\usepackage{cleveref}
\usepackage{flushend}
\bibliography{references/main}

\newcommand{\sys}{\textsc{Jwins}\xspace}
\newcommand{\FL}{\ac{FL}\xspace}
\newcommand{\DL}{\ac{DL}\xspace}

\newcommand{\topk}{\textsc{TopK}\xspace}
\newcommand{\femnist}{FEMNIST\xspace}
\newcommand{\celeba}{CelebA\xspace}
\newcommand{\shakespeare}{Shakespeare\xspace}
\newcommand{\cifar}{CIFAR-10\xspace}
\newcommand{\movielens}{MovieLens\xspace}
\newcommand{\sgd}{{\xspace}\ac{SGD}\xspace}
\newcommand{\dpsgd}{{\xspace}\ac{D-PSGD}\xspace}
\newcommand{\iid}{\ac{IID}\xspace}
\newcommand{\niid}{\ac{non-IID}\xspace}
\newcommand{\fullshare}{{\xspace}full-shar\-ing\xspace}
\newcommand{\subsampling}{{\xspace}random sampling\xspace}
\newcommand{\xsym}{\boldsymbol{x}\xspace} \graphicspath{ {figures/} }
\usepackage{tikz}
\usepackage[eulergreek]{sansmath}

\usepackage{acronym}
\acrodef{DL}{decentralized learning}
\acrodef{ML}{machine learning}
\acrodef{D-PSGD}{decentralized parallel stochastic gradient descent}
\acrodef{FL}{federated learning}
\acrodef{SGD}{stochastic gradient descent}
\acrodef{IID}{independent and identically distributed}
\acrodef{non-IID}{non independent and identically distributed}
\acrodef{RMSE}{root mean square error}
\acrodef{RMW}{random model walk}
\acrodef{GL}{gossip learning}
\acrodef{DWT}{discrete wavelet transform}
\acrodef{FFT}{fast Fourier transform} %

\newcommand\copyrighttext{%
	\footnotesize \textcopyright 2023 IEEE.
	Personal use of this material is permitted.
	Permission from IEEE must be obtained for all other uses,
	in any current or future media, including reprinting/republishing this
	material for advertising or promotional purposes, creating new collective
	works, for resale or redistribution to servers or
	lists, or reuse of any copyrighted component of this work in other works.
	Pre-print version. Presented in the {43rd IEEE International Conference on Distributed Computing Systems (ICDCS '23)}. For the final published version, please refer to DOI \href{https://doi.org/10.1109/ICDCS57875.2023.00067}{10.1109/ICDCS57875.2023.00067}.}
\newcommand\copyrightnotice{%
	\begin{tikzpicture}[remember picture,overlay]
		\node[anchor=south,yshift=10pt,fill=yellow!20] at (current page.south) {\fbox{\parbox{\dimexpr\textwidth-\fboxsep-\fboxrule\relax}{\copyrighttext}}};
	\end{tikzpicture}%
}

\begin{document}

\title{Get More for Less in \\Decentralized Learning Systems}

\author{\IEEEauthorblockN{Akash Dhasade\IEEEauthorrefmark{1}, Anne-Marie Kermarrec\IEEEauthorrefmark{1}, Rafael Pires\IEEEauthorrefmark{1}, Rishi Sharma\IEEEauthorrefmark{1}\IEEEauthorrefmark{2}, Milos Vujasinovic\IEEEauthorrefmark{1} and Jeffrey Wigger\IEEEauthorrefmark{3}}
\IEEEauthorblockA{\IEEEauthorrefmark{1}\emph{EPFL, Switzerland}, \texttt{first.last}@epfl.ch}
\IEEEauthorblockA{\IEEEauthorrefmark{2}\emph{Corresponding author}. \IEEEauthorrefmark{3}\emph{Unaffiliated}.}
}

\maketitle
\copyrightnotice

\pagestyle{plain}

\begin{abstract}
	\Ac{DL} systems have been gaining popularity because they avoid raw data sharing by communicating only model parameters, hence preserving data confidentiality.
	However, the large size of deep neural networks poses a significant challenge for decentralized training, since each node needs to exchange gigabytes of data, overloading the network.
	In this paper, we address this challenge with \sys, a communication-efficient and fully decentralized learning system that shares only a subset of parameters through sparsification.
	\sys uses wavelet transform to limit the information loss due to sparsification and a randomized communication cut-off that reduces communication usage without damaging the performance of trained models.
	We demonstrate empirically with 96 \ac{DL} nodes on non-IID datasets that \sys can achieve similar accuracies to full-sharing \ac{DL} while sending up to 64\% fewer bytes.
    Additionally, on low communication budgets, \sys outperforms the state-of-the-art communication-efficient \ac{DL} algorithm \textsc{Choco-SGD} by up to 4\texttimes~in terms of network savings and time.
\end{abstract}
 

\acresetall
\section{Introduction}

Deep learning algorithms have significantly improved artificial intelligence tasks, such as image classification~\cite{he2015delving}, speech recognition~\cite{hannun2014deep}, and text detection~\cite{8237349}.
Enormous volumes of data are produced daily on smartphones and edge devices, which greatly enhance the performance of deep learning models. 
In addition to facing challenges in efficiently handling the massive size of users' data, the centralized data collection also poses a considerable privacy threat, since personal data can be sensitive to users and its exposure represents a significant concern. 

\Ac{DL} has emerged as an attractive alternative for training models on decentralized data, where nodes learn by exchanging models among themselves instead of sharing data with large service providers. 
This approach provides additional privacy~\cite{lian2017dpsgd, pmlr-v119-koloskova20a,koloskova2019decentralized, koloskova2020decentralized} and obviates the need for complex data management in data centers.
Furthermore, \ac{DL} systems are also popular for their enhanced scalability compared to centralized systems, where the server can become the bottleneck of the system~\cite{lian2017dpsgd}. 
Nodes in \ac{DL} are typically devices at the edge of the network, with limited bandwidth, and connected according to a communication topology.
While powerful in computational capabilities, these devices must exchange \ac{ML} models under constrained communication budgets.
As a consequence, minimizing the number of bytes transferred when performing \ac{DL} training has been a major focus of research~\cite{tang2018communication, koloskova2019decentralized, koloskova2020decentralized, vogels2020practical}.

An important technique to minimize the communication during training is \emph{sparsification} or \textit{partial sharing}, which limits the model transfer to a small fraction of \emph{important} parameters~\cite{7447103, strom2015scalable, alistarh2018sparseconvergence, lin2018deep}.
For sparsification to be effective, it is crucial to select these important parameters with care, ensuring that the resulting global model remains unbiased and maintains the expected accuracy.
The rest of the parameters remain local in favor of communication efficiency, thus limiting the information exchanged between nodes.
Considering that the data distribution across nodes is often \niid, the withheld information due to sparsification can have a detrimental effect on model utility~\cite{hsiehskewscout2020}.
While there have been in-depth theoretical studies in sparsification-based communication efficiency for \ac{DL} systems~\cite{koloskova2019decentralized, koloskova2020decentralized}, empirical studies with hundreds of nodes and challenging \niid datasets have been fairly limited.
Most settings either use a central coordinator or all-to-all communication~\cite{hsiehskewscout2020, hsieh2017gaia, 7447103}, which are both unpractical in decentralized systems.

In this paper, we present \sys (\emph{just what is needed sharing}), a system that performs decentralized training under limited communication budgets while retaining good model performance.
The novelty of \sys lies in the way \sys addresses the critical issue of choosing important parameters by (1) ranking the parameters and averaging the sparse model vectors in the wavelet-frequency domain, and (2) choosing the size of the sparse model vector using a randomized cut-off scheme which respects the overall communication budget.
Since each wavelet coefficient contains information about a subset of the original parameters, \sys can pack more information in the sparse vectors compared to standard sparsification techniques that rely only on the magnitude of parameter updates.
The parameters are both shared and averaged as wavelet coefficients, thus limiting the loss of information due to parameter sparsification in each round.
Finally, with the randomized cut-off scheme, each node gets the chance to share the full model parameters every few rounds without violating the communication budget of the entire training process.
This shift of approach to the wavelet-frequency domain results in scalable and performant sparsification.

We evaluated \sys in \niid data distributions over the extensively used datasets of \cifar, \movielens, and the standard LEAF benchmarks of \shakespeare, \celeba and \femnist~\cite{krizhevsky2014cifar, movielensdataset2015, leaf}. 
Our results for a \num{96}-node training demonstrate that \sys can achieve an effective sparsification of parameters, reducing the model to $36\%$ of its original size, mostly without impact on test accuracy.
In the worst case, \sys causes an accuracy drop of no more than $3\%$ when compared to full-sharing decentralized learning.
In absolute terms, this translates to reductions in network traffic of up to \SI{1.3}{\tebi\byte} against \fullshare.
Our scalability study of up to \num{384} nodes demonstrates that \sys scales gracefully with increasing number of nodes.
To reach the same testing accuracy given a communication budget, \sys outperforms the sparse communication baseline of \subsampling and the state-of-the-art communication compression algorithm called \textsc{Choco-SGD}~\cite{koloskova2019decentralized} by up to $4\times$ in terms of communication rounds and wall-clock time.
Finally, given the exact same communication budget, \sys reaches 2-10\% better accuracies %
against \subsampling sparsification and \textsc{Choco-SGD}.  %

In this paper, we make the following contributions:
\begin{itemize}
	\item We propose the design of \sys: a novel decentralized learning algorithm that limits the number of parameters exchanged during the training, building on top of basic sparsification. \sys relies on a smart combination of wavelet transform and randomized cut-off strategy to limit communication without affecting the accuracy of the learning task.
	\item We empirically demonstrate that model sparsification and averaging in the wavelet-frequency domain are highly effective in decentralized learning. 
        To the best of our knowledge, we are the first to apply wavelet transform for sparsification and averaging in \ac{DL} systems.
	\item We conducted an extensive evaluation study on several learning tasks with \niid data distributions, including image classification, recommendation, and next character prediction.
        Our experimental results demonstrate that \sys outperforms \subsampling and \textsc{Choco-SGD}, and easily scales up to 384 nodes.
\end{itemize}

\section{Background}
\label{sec:background}
We briefly describe the \DL setup and overview a few communication compression techniques to set the stage for \sys  in Section~\ref{sec:partialModel}. %

\subsection{Decentralized learning}
\label{subsec:background_DL}
\paragraph{Objective.} The decentralized learning setting comprises $n$ nodes seeking to jointly learn an ML model.
Each node $i \in [n]$ has access to samples from the local data distribution $D_i$ which typically varies across nodes.
The goal of learning is to find the parameters of the model $\xsym$ that perform well on the union of datasets spread across nodes by minimizing the average loss, as follows: 
\begin{equation}
	\label{eqn:DL_obj}
	\min_{\xsym \in \mathbb{R}^d} \left[f(\xsym) =\frac{1}{n} \sum_{i = 1}^n  \underbrace{\mathbb{E}_{\xi \sim D_i}[F_i(\xsym; \xi)]}_{=:f_i(\xsym)}\right]
\end{equation}
where $f_i(x)$ refers to the local objective function of node $i$ and $F_i(\xsym; \xi)$ corresponds to the (possibly non-convex) prediction loss on sample $\xi$ under the model parameters $\xsym$.
The local objectives $f_i$ are evaluated as the expected value of loss $F_i$ on the local data distribution $D_i$.
Finally, the nodes in \DL are connected through a network topology $G = (V,E)$ which defines the neighborhood of nodes for communication. 
Every $\{i, j\} \in E$ corresponds to an undirected edge between nodes $i$ and $j$.

\paragraph{Decentralized training.} 
\DL training occurs in repetitions of training rounds.
In each round, the nodes follow a general \textit{train-communicate-aggregate} pattern.
The training typically consists of running several steps of an optimization algorithm like \sgd on mini-batches of data sampled from the local data distribution $D_i$, which is then shared with all (\eg, \dpsgd~\cite{lian2017dpsgd}) or a subset of neighbors (\eg, \ac{RMW}~\cite{ormandi:2013:gossiplearning}).
Finally, in the aggregation stage, nodes combine the received models with their own, either by performing a plain (\ac{RMW}) or weighted averaging (\dpsgd). 
We highlight that \sys concerns only the \textit{communication} stage in \DL, and it is independent of the specific \emph{aggregation} algorithm. %

\subsection{Communication compression}
\label{sec:compression}

Apart from more traditional and general-purpose compression algorithms like LZ4~\cite{collet:lz4:2022} and LZMA~\cite{pavlov:lzma:2022}, model quantization~\cite{seide2014-bit, alistarh2017quantization} and sparsification~\cite{7447103, strom2015scalable, lin2018deep} are also referred to as compression techniques in ML.
Given that model parameters are floating point numbers, quantization aims at representing such numbers as integers with a smaller bit length. %
Sparsification, in turn, focuses on reducing the number of parameters or gradients actually shared. 
\sys falls into this sparsification category, as we will detail later (\Cref{sec:partialModel}).  %

\subsubsection{Gradient sparsification.}
\label{subsec:grad_sparsification}

Having a synchronized model across all worker nodes on every training round, architectures with a central coordinator such as parameter server~\cite{li:2014:parameterserver} and \FL~\cite{mcmahan2017communicationefficient} can afford to share and aggregate gradients rather than parameters.
The reason is that worker nodes depart from the same point in the parameter space, and therefore their resulting computed gradients have a common reference point.
In such case, sparsification can be trivially done by taking the gradients with the largest magnitude, \ie, all that fall beyond a given threshold.
Once the server receives the gradients from workers, it suffices to apply these gradients to the global model before the next training iteration.
This method, known as \topk~\cite{alistarh2018sparseconvergence}, sets the importance of a parameter by the magnitude of the corresponding gradient in a given round.

\paragraph{Sparsification and metadata.}
Whenever the gradients for the full model are shared, they can be directly aggregated to the corresponding model parameter according to their position in a serialized vector.
Sparsified gradients, on the other hand, do not hold such a positional correspondence to the original model.
Hence, one must also send a list of parameter indexes with which the sparsified gradients are associated.
This additional metadata incurs supplementary costs in network usage.

\paragraph{Accumulation.}
Depending on the model architecture, different parameters may incur changes of distinct magnitudes over the training.
For a high degree of sparsification, the gradients corresponding to the same parameters may end up being shared all the time, hence neglecting parts of the model and producing lower-quality models.
To avoid this, the gradients $g_i^{(t)}$ of node $i$ at the round $t$ are first accumulated into the residual vector $u_i^{(t)}$~\cite{seide2014-bit, aji-heafield-2017-sparse}, as shown below, where $\eta$ is the learning rate.
\begin{equation} \label{eq:gradientAccumulation}
	u_i^{(t)} = u_i^{(t-1)} - \eta \cdot g_i^{(t)}
\end{equation}
Subsequently, \topk is applied on this residual vector, which allows parameters with smaller gradients to be shared at some later point of time.
The residual values $u_i^{(t)}$ of shared gradients are reset to zero every time they are selected for sharing.

\subsubsection{Parameter sparsification.}
In contrast to the central-coordinator architectures, there is no central entity for synchronizing models in the decentralized learning setting.
Nodes in \DL may have distinct models at the beginning of each round, and their computed gradients are incompatible for direct aggregation since these gradients represent the local progress from different points in the parameter space.
As a consequence, instead of gradients, in \DL the final model parameters must be shared and aggregated.
This prevents us from directly applying gradient sparsification techniques in \ac{DL}. %

\paragraph{Random sampling.}
\topk sparsification and accumulation from gradient sparsification can be adapted to \ac{DL} settings for parameter sparsification, but are inefficient~\cite{tang2020sparsification}.
Instead, \subsampling is often used~\cite{koloskova2019decentralized, tang2020sparsification, HEGEDUS2021109} where a random subset of the parameters of a predefined size is chosen and shared.
Further, when a common pseudo-random generator is used across collaborating nodes, \subsampling brings along the benefit of drastically reducing the sparsification metadata.
The reason is that just sharing the seed (an integer) from which the random indices are generated suffices for mapping them back to their corresponding model parameters.

\section{\sys}
\label{sec:partialModel}

\begin{figure}
	\centering
	\includegraphics[width=0.7\linewidth]{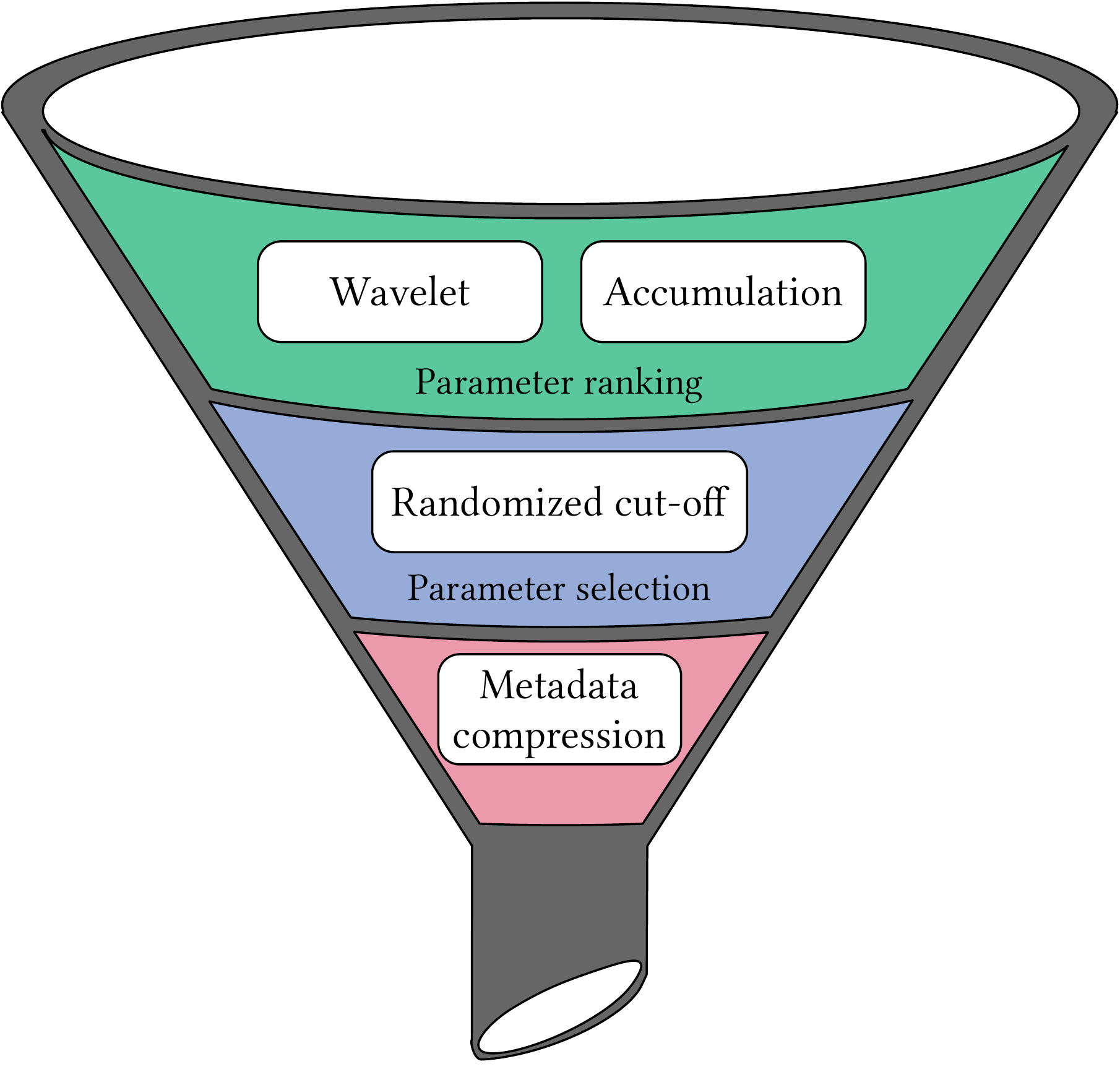}
	\caption{\sys consists of four main modules that produce a smaller partial model: \emph{(i)} wavelet transform and \emph{(ii)} accumulation gives importance scores to parameters; \emph{(iii)} randomized cut-off enables nodes to randomly choose the fraction of shared parameters; and \emph{(iv)} metadata compression is used to practically nullify the overheads of metadata when sharing sparsified models.}
	\label{fig:jwins}
\end{figure}

In this section, we describe \sys and its components in detail. \sys is built using four core modules.
The first one is the representation in the wavelet frequency domain of the model and the model change. %
The second module is the accumulation of these changes into an importance score for the parameters.
These first two modules are outlined in \Cref{sec:waveletAccumulation}, as part of the \sys parameter ranking. 
\Cref{sec:randomized} then brings the third module by explaining how \sys selects the best-ranked parameters.
The fourth module treats how \sys compresses the metadata of the sparsified model generated for sharing and is covered in \Cref{sec:metaCompression}. 
These four core modules are illustrated in \Cref{fig:jwins}.
Finally, in \Cref{alltogether}, we present a summary of the \sys operation, detailing how information circulates through the modules and the process of model averaging in the wavelet domain.

In the sections that follow, we use subscripts to index nodes and superscripts to index global rounds numbers and local steps. 
Therefore $x^{(t,s)}_i$ refers to the model in $t-$th global round after $s$ local steps at node $i$.
Further, in each round $t \in [T]$, nodes perform a total of $\tau$ local steps \ie $s \in [\tau]$.

\subsection{\sys parameter ranking} \label{sec:waveletAccumulation}

\sys parameter ranking generates and maintains a ranking order of the model parameters and relies on the combination of wavelet transform and accumulation.
To the best of our knowledge, we are the first to apply them together in decentralized learning systems.

\paragraph{Parameter and gradient representation.}

\begin{figure}[t]
	\centering
	\includegraphics{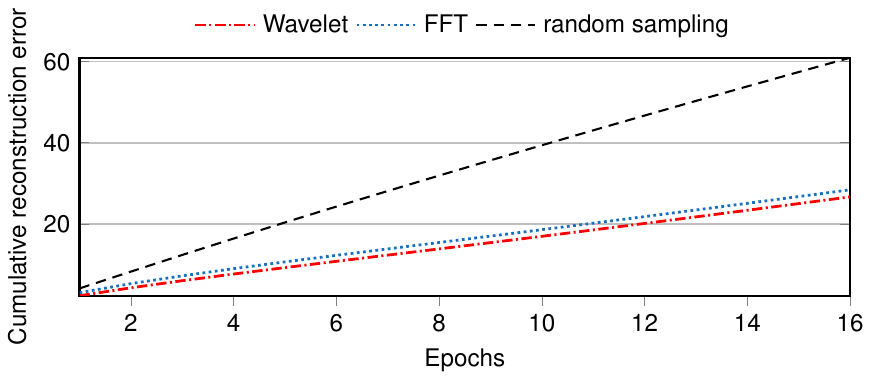}
	\caption{Mean squared error between the original and reconstructed model when sparsifying parameters using the given algorithms. The plot exhibits the information loss due to sparsification.}
	\label{fig:wavelet-justify}
\end{figure}

\Ac{DWT} and \ac{FFT} have been used commonly for image compression~\cite{mantoro2017comparison, ramkumar1997fft, patel2016improved}.
For instance, JPEG2000 standard~\cite{taubman2002jpeg2000} uses the properties of \ac{DWT} for compression.
Deriving the inspiration from image compression, we test the performance of compression techniques based on \ac{DWT} and \ac{FFT} for neural networks in the presence of sparsification on \cifar~\cite{krizhevsky2014cifar} using GN-LeNet~\cite{hsiehskewscout2020} given a communication budget of \num{10}\%.
\Cref{fig:wavelet-justify} presents the cumulative reconstruction error of \ac{DWT} and \ac{FFT} against \subsampling sparsification for a single node training.
This experiment simulates an exchange of model parameters where only a sparsified model is received by applying parameter sparsification in the transformed domain.
The reconstruction error is computed as the mean squared error (MSE) of the model without any compression for this round against this sparsified model. %
Objectively, the algorithm with the lowest cumulative reconstruction error loses the least information on sparsification.
Being the compression technique with the least cumulative reconstruction error, we incorporate DWT in \sys.

In \sys, we use a four-level discrete wavelet decomposition with Symlet2 (Sym2) wavelets to represent the model $x_i^{(t,s)}$ in the wavelet domain. %
We experimented with different wavelet functions and Sym2 outperformed the others.
Increasing the levels beyond four did not have any noticeable improvements.
Intuitively, this representation when compared to the original parameter domain has the advantage of containing coefficients representing the model at lower wavelet frequencies, i.e., single coefficients at higher levels contain information about a neighborhood of parameters in the original domain. 
Therefore, a sparse wavelet coefficient vector with $K$ non-zero elements packs more information about a change in the model than a vector with $K$ parameters in the original domain.

\paragraph{Accumulation in the wavelet domain.}
In order to decide which wavelet coefficients should be shared with neighboring nodes, we adapted to the decentralized learning setting the techniques of gradient sparsification and accumulation, as described in \Cref{subsec:grad_sparsification}.
Instead of gradients, \sys tracks model parameter changes during training and transforms them to the wavelet frequency domain for accumulation. The following equation generates the score:
\begin{equation}
	\label{eq:ranking}
	V_i^{\prime(t)} = V_i^{(t)} + \textsc{DWT}(x_i^{(t,\tau)} - x_i^{(t, 0)})
\end{equation}  %
$V_i^{\prime(t)}$ represents the new importance score of the model parameters, which is equal to the previously accumulated score $V_i^{(t)}$ added to the difference of the parameters in the wavelet frequency domain over a local training of $\tau$ SGD steps. 
\sys, however, does not restrict the use of SGD or stateless optimizers. We only use SGD for simplicity.
This score from \Cref{eq:ranking} is used by the \sys parameter selection to share the important parameters.
The elements of $V_i^{(t)}$ that correspond to the chosen parameters to be shared are set to zero .
The final accumulation step at the end of the round is captured in the equation below:
\begin{equation}
	\label{eq:accumulation}
	V_i^{(t+1)} = V_i^{(t)} + \textsc{DWT}(x_i^{(t+1,0)} - x_i^{(t, 0)})
\end{equation}
where $x_i^{(t+1,0)}$ is the model after the averaging.

\subsection{\sys parameter selection} \label{sec:randomized}
\sys parameter selection uses \topk sparsification on the absolute value of the importance scores generated by \sys parameter ranking.
The size of this selection is governed by a randomized cut-off that determines the subset of model parameters being shared.
As noted in \Cref{subsec:grad_sparsification}, simple \topk-based sparsification causes the same set of parameters getting repeatedly shared, in detriment of lower-ranked parameters. 
While this is partly fixed by the parameter ranking based on wavelet and accumulation, \sys further attenuates this negative effect by sharing random proportions of the model in every round.

To select the number of best-ranked model parameters to be shared, \sys randomized cut-off component independently chooses in each node a random sharing percentage $\alpha$ from a predetermined distribution based on the available communication budget.
Its benefits are threefold: \emph{(i)} parameters that change too slowly can be picked before reaching the threshold; \emph{(ii)} it prevents all nodes from congesting the network by using a large $\alpha$ at the same time; and \emph{(iii)} it prevents a collective drop in model quality due to herd behavior.
The negative effect of herd behavior happens when nodes collectively move to a larger $\alpha$, since they suddenly start sharing parameters that were over-specialized in each individual local data. 
Averaging these over-specialized parameters together has a similar effect to that of resetting the learning for those parameters, which induces losses in performance.
To sum up, several problems arise by having a global sharing fraction $\alpha$: \emph{(i)} a small $\alpha$ causes the oblivion of slow-changing parameters; \emph{(ii)} large $\alpha$ causes network congestion; and \emph{(iii)} instant global changes in $\alpha$ causes a drop in the quality of the trained model.
Put simply, our randomized cut-off strategy avoids them all.

\subsection{Metadata compression} \label{sec:metaCompression}
The metadata compression module is a crucial component of \sys, as it plays a vital role in reducing the amount of data transferred. 
Without compression and optimization, the metadata from parameter selection would effectively double the number of bytes transmitted, as each parameter is associated with an index.
Data compression in \DL typically uses floating point compression to perform parameter compression (see \Cref{subsec:exp_setup} for \sys parameter compression). %
However, in the case of indices, which are integers, we conducted experiments using various general-purpose compression algorithms.
After careful evaluation, we determined that applying Elias gamma~\cite{elias1975universal} compression to the difference array of indices, similar to the approach used in \textsc{QSGD}~\cite{alistarh2017quantization}, yielded the best compression rate with a minimal time overhead.

\subsection{\sys at work} \label{alltogether}
\begin{algorithm2e}
	\DontPrintSemicolon
	\caption{\sys on Node i} \label{alg:jwins}
	\KwIn{Initialize $x^{(0,0)}_i=x^{(0,0)}$ and $V_i={}\pmb{0}$; 
		the weight matrix $W$; metadata compression algorithm $C$; local dataset $D_i$;
		the learning rate $\eta$; the number of communication steps $T$; the set of neighbours $N_i$; %
    and the number of local SGD steps $\tau$} %
	\KwOut{The trained model $x^{(T, 0)}_i$}
	\For{$t \gets 0,1,\ldots T$}{ %
		\For{$s \gets 1$ to $\tau$}{ %
			Draw a random mini-batch $\xi_i$ from $D_i$\;
			$x^{(t,s)}_i \gets x^{(t,s-1)}_i - \eta \cdot \nabla F_i(x^{(t,s-1)}_i; \xi_i)$\; 
		}
            Accumulate $\textsc{DWT}(x^{(t,\tau)}_i - x^{(t,0)}_i)$ in $V'_i$ \hfill$\triangleright $ {\color{purple}\cref{eq:ranking}}\;
            Randomly choose the communication budget $K_i^{(t)}$\;
            Get $\textsc{TopK}_i^{(t)}$ indices $I_i$ from accumulated $V'_i$\; 
            Send $\textsc{DWT}(x^{(t,\tau)}_i)[I_i]$ and $C(I_i)$ to all $j \in N_i$\;
            Receive partial wavelets $Z^{(t,\tau)}_j$ from all $j \in N_i$\;
            Average all received $Z^{(t,\tau)}_j$ with $\textsc{DWT}(x^{(t,\tau)}_i)$ \;
            Get $x^{(t+1,0)}_i$ by applying $\textsc{DWT}^{-1}$ on the average \;
            Reset $V_i$ for sent and updated parameters \hfill$\triangleright $ {\color{purple}\cref{eq:accumulation}}\;
}
\end{algorithm2e} We now describe how the various components of \sys work together.
A full pseudocode of \sys, executed concurrently on each node $i$ is given in \Cref{alg:jwins}. 
At the start of round $t$, the node performs $\tau$ training steps on its local data, resulting in the model $x_i^{(t, \tau)}$.
The parameter ranking component calculates the model difference, $\Delta x = (x_i^{(t, \tau)} - x_i^{(t, 0)})$. This difference is then transformed using DWT and added to an accumulation vector ($V$) to update rankings for the current round.
\sys parameter selection determines the value of $\alpha$ for the round based on a randomized cut-off, which determines the percentage of parameters to share.
From the accumulation vector, the top $K_i^{(t)}$ indices ($I_i$) are selected. 
The corresponding coefficients in $DWT(x_i^{(t, \tau)})$ for these indices, along with $C(I_i)$, the compressed indices, are shared with neighboring nodes in the topology.
The received sparse set of wavelet coefficients from all neighbors are averaged using a predefined weighting scheme. The resulting averaged coefficients are then inverted using $DWT^{-1}$ to obtain $x_i^{(t+1, 0)}$, \ie, the updated model for the next round.
Entries in the accumulation vector ($V_i$) that were chosen in this round are set to zero.
Finally, $V_i$ is updated to account for the model change due to averaging, and the round is concluded.

\section{Evaluation}
\label{sec:eval}

We present here an extensive evaluation of \sys in a decentralized setting.
We start by describing \sys implementation and the experimental setup.
Then we present the metrics and results across various experiments.
Finally, we discuss the implications of the performance of \sys. %

\subsection{Implementation}
\sys is written in only \num{800} lines of Python 3.8 code on top of the DecentralizePy framework~\cite{dhasade:2023:dcpy}\footnote{Source code is available at \url{https://github.com/sacs-epfl/decentralizepy}.}.
\sys uses PyTorch v1.10.2~\cite{torchSoftware} for ML along with the {\small \texttt{torch.multiprocessing}} for creating processes.
Each process represents a real \ac{DL} node with its own data and is independent of other processes.
ZeroMQ~\cite{boccassi:2022:zmq} is used to communicate between processes via TCP sockets.
The discrete wavelet transformations of the model weights are calculated using the open-source PyWavelets library~\cite{Lee2019}.
We use a common seed for all pseudo-random generators in a node for reproducibility.
\sys is modular, easily extensible, and can support new neural network architectures, datasets, topologies, and compression techniques by plugging other modules in.

\subsection{Experimental setup}
\label{subsec:exp_setup}

\paragraph{Cluster deployment and network}
Our experimental set\-up comprises 6 machines, each equipped with 2 Intel(R) Xeon(R) CPU E5-2630 v3 @ 2.40GHz of 8 cores. Hyperthreading is enabled and the machines run Ubuntu 20.04.4 LTS with 5.4.0-99-generic kernel.
For each experiment, we run 96 \ac{DL} nodes (processes) equally distributed across the machines, connected in a random $d$-regular topology with $d = 4$, \ie the degree of all vertices is equal to $d$.
For the scalability experiments (\Cref{{sec:scalability}}), we run up to 384 \ac{DL} nodes. 

\paragraph{Learning tasks and hyperparameters}
We evaluate \sys over \niid train and test datasets of \cifar~\cite{krizhevsky2014cifar}, \movielens~\cite{movielensdataset2015} and the standard \ac{FL} LEAF benchmarks~\cite{leaf} of \celeba, \femnist, and \shakespeare.
These datasets are widely used across related works~\cite{dhasade:2022:rex, koloskova2020decentralized, mcmahan2017communicationefficient, tang2020sparsification}.
The image classification task for the first three datasets is achieved over convolutional neural networks of varying specifications~\cite{leaf, hsiehskewscout2020}.
Recommendation over \movielens is performed through matrix factorization~\cite{korenmatrixfactorization2009} using embeddings.
Finally, the next-character prediction for \shakespeare uses a stacked LSTM~\cite{leaf}.
This variety of learning tasks and models is chosen to demonstrate the generality of \sys.
The hyperparameters of learning rate ($\eta$), number of communication rounds per epoch (r), and batch size (b) are tuned by running a grid search for the full-sharing baseline.
This is done separately on data unseen during training.
The same values of hyperparameters are then used for \sys and other baselines.
The basic \sgd optimizer without momentum is used.
The decentralized learning algorithm is \dpsgd over Metropolis-Hastings weights~\cite{XIAO200465}.

\paragraph{Baselines} 
We start by comparing \sys against two decentralized system baselines: \textit{(i)} where the whole set of parameters is exchanged during each communication round (called \emph{\fullshare} in the rest of the paper), and \textit{(ii)} with random sampling sparsification.
Among the sparsification algorithms of \subsampling and \topk, we only present \subsampling because \topk overfits to local data as seen in the ablation study (\Cref{sec:breakingjwins}) for \sys without wavelet (essentially, \topk).
The comparison with \fullshare is meant to assess the performance of \sys with respect to the resulting accuracy while \subsampling is used as a baseline for network savings.
Next, we compare \sys against \textsc{Choco-SGD}~\cite{koloskova2019decentralized}, the state-of-the-art algorithm for communication-efficient decentralized learning with \topk as the compression technique (referred to as \textsc{Choco} hereafter).
\textsc{PowerGossip}~\cite{vogels2020practical} is another strong communication-efficient algorithm for \ac{DL}, but it performs as good as tuned \textsc{Choco} in their experiments.
Hence, we only compare against \textsc{Choco} here.

\paragraph{Data partitioning}
For the image classification task of \cifar, the dataset is sorted by labels and partitioned into $2n$ shards, where $n$ is the number of nodes.
Each node receives 2 shards randomly picked, effectively limiting the maximum number of classes per node to 4, which makes the data distribution \niid~\cite{mcmahan2017communicationefficient}.
All other datasets contain a mapping between clients and their data samples. 
The term \emph{clients} here refers to the users who originally produced a data sample, whereas  \emph{nodes} pertains to processes representing a computing unit in our empirical test-bed.
For example, the data samples in \femnist are grouped by the clients who wrote the digits, and the data samples in \movielens are grouped by the clients who produced an item-rating pair.

We distribute the data produced by these clients over the 96 training nodes such that each node receives an equal number of clients.
Since the \shakespeare dataset contains a very large number of samples, we only distribute 96 out of 660 clients from the original dataset (called \shakespeare onwards), each containing between \num{800} and \num{1250} samples.
We do this in order to limit the training time and to homogenize the computing requirement per node.
This however causes a drop in the reached accuracy, which we consider acceptable for the sake of comparison to \fullshare and \subsampling under the same setup.

\paragraph{Compression}
Due to the large number of model parameters, we empirically assessed multiple compression algorithms in order to reduce the amount of data for transferring these models, and therefore minimize communication.
We chose Fpzip compression algorithm~\cite{lindstrom2006fpzip} since it performed the best across our experiments.
This compression is applied uniformly for all the model parameters and for all experiments and baselines.
For metadata compression, we use Elias gamma~\cite{elias1975universal} as discussed in \Cref{sec:metaCompression}.

\begin{table*}[t!]
	\small
	\centering
	\caption{\label{tab:performanceBasic}Final test accuracies and network transfers for \sys in comparison to \fullshare and \subsampling for plots in Figure~\ref{fig:sysVsDPSGDVsSubsampling}. 
	\sys achieves as good test accuracy as \fullshare across most datasets,  while it is 2\%-15\% better than \subsampling.
	\sys saves several hundreds of gigabytes in network transfers.
	} 
		\begin{tabular}{ c r c c c r r r }
			\toprule
			\multirow{2}{*}{\textsc{Dataset}} & \multirow{2}{*}{\textsc{Epochs}} & \multicolumn{3}{@{} c}{\textsc{Test accuracies of algorithms} $\uparrow$} & \multicolumn{2}{@{} c}{\textsc{Total data sent [\SI{}{\gibi\byte}]} $\downarrow$} & \multirow{2}{*}{\textsc{Network savings}}\\
			& & \fullshare & \subsampling & \sys & \fullshare & \sys & \\ 
			\toprule
			\cifar & \num{268} & \num{58.3} & \num{40.1} & \num{55.3} & \num{628.2} & \num{231.2} &  \num{62.2}\%  \\ 
			\movielens & \num{400} & \num{91.7} & \num{89.1} & \num{92.6} & \num{1103.5} & \num{394.6} & \num{64.2}\% \\ 
			\shakespeare & \num{57} & \num{35.0} & \num{30.5} & \num{34.5} & \num{2127.2} & \num{753.7} & \num{64.6}\% \\
			
			\celeba & \num{26} & \num{89.7} & \num{89.0} & \num{90.9} & \num{10.4} & \num{3.8} & \num{63.5}\%  \\ 
            \femnist  & \num{25} & \num{80.6} & \num{79.6} & \num{81.6} & \num{557.5} & \num{199.2} & \num{64.3}\% \\ 
            \bottomrule
	\end{tabular}%
\end{table*}
 
\begin{figure}
	\centering
	\includegraphics{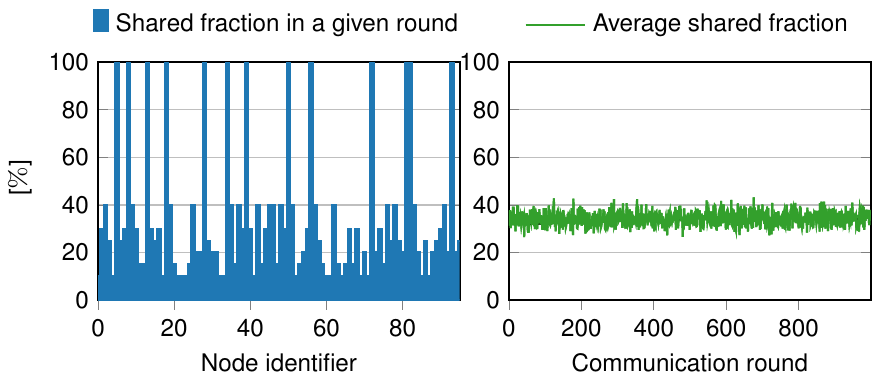}
	\caption{Randomized cut-off in \sys. 
		Chart on the left depicts the random percentages selected by \sys' nodes in a typical communication round.
		Chart on the right shows the average sharing percentage across nodes over communication rounds.} %
\label{fig:data_alphas}
\end{figure}

\begin{figure*}[h]
	\centering
	\includegraphics{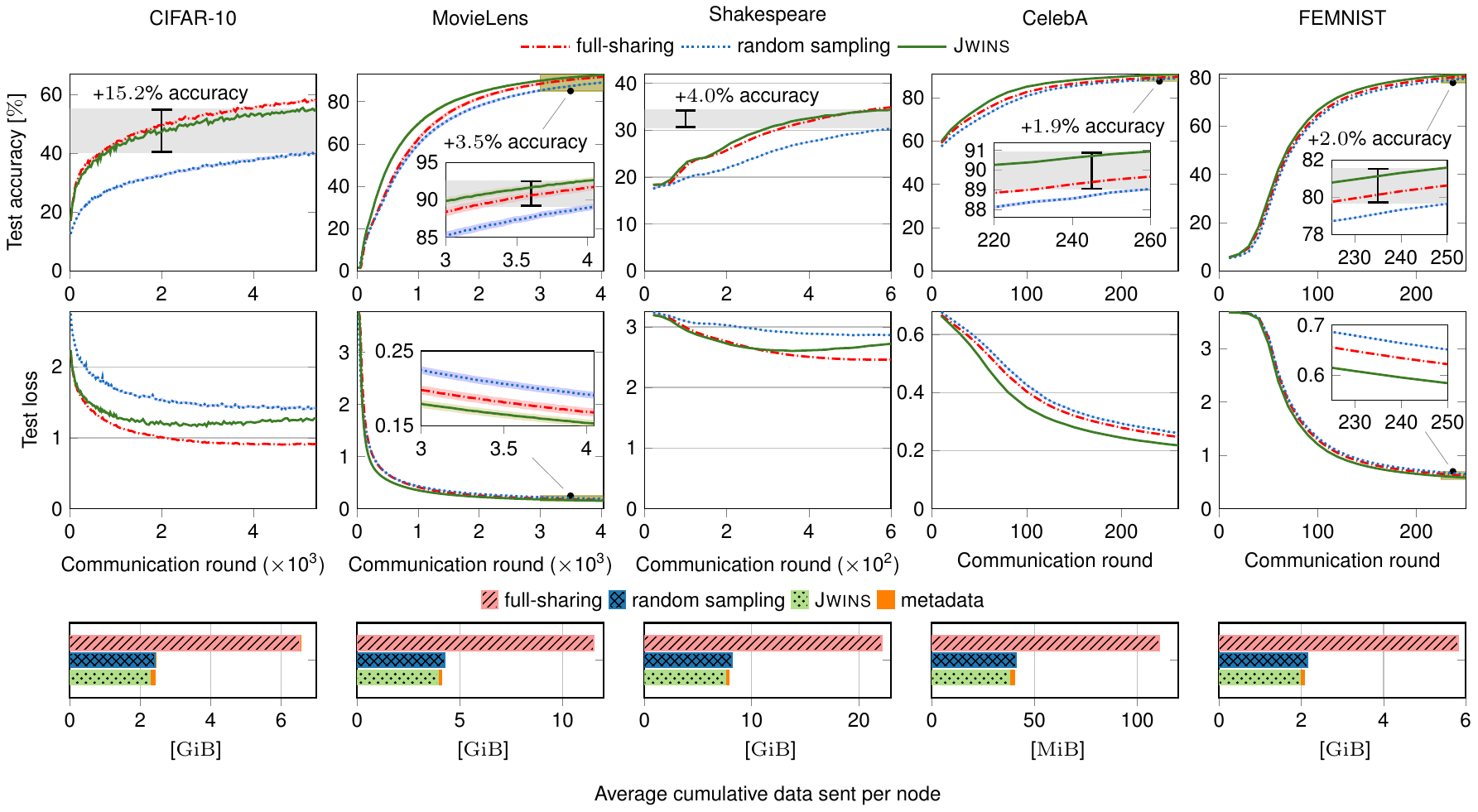}
	\caption{Learning curves and network usage for \sys compared to \fullshare and \subsampling when run for fixed rounds. 
		\sys achieves as good test accuracy and test loss as \fullshare across most datasets  (row-1 and row-2), while requiring significantly less network transfers per node (row-3). 
		Results are further quantified in \Cref{tab:performanceBasic}.}
	\label{fig:sysVsDPSGDVsSubsampling}
\end{figure*}

\paragraph{Communication}
The communication load for \fullshare per iteration is fixed as the full model is always exchanged.
In contrast, the communication load for \sys is randomized but governed by the $\alpha$ distribution used by \sys parameter selection.
In \sys, $\alpha \in \{10\%, 15\%, 20\%, 25\%, 30\%, 40\%, 100\%\}$, \ie $\alpha$ is uniformly randomly picked from this list every round by every node independently.
Empirically, this works well for all the tasks with minimal accuracy trade-offs.
\Cref{fig:data_alphas} shows this distribution in action in a \sys run.
For a fair comparison of the learning performance of \sys and \subsampling, we limit the communication budget per communication round of \subsampling to match the total communication by \sys.
This results in a sharing of 37\% of model parameters in every iteration of \subsampling. %
We also evaluate \sys in settings with lower communication budgets of $20\%$ and $10\%$ against \textsc{Choco}.

\paragraph{Metrics}
We evaluate \sys along the following metrics: (1) average top-1 test accuracy of nodes on the test set, (2) average loss of nodes on the test set, (3) average number of bytes transferred by nodes.
These are real values measured by instrumenting the experiments.
Every experiment is conducted five times, using different random seeds that determine the data distribution, neighbors, and initial weights. 
The values we provide are averages from these five runs, within a 95\% confidence interval.

\begin{figure*}[t!]
	\centering
	\includegraphics{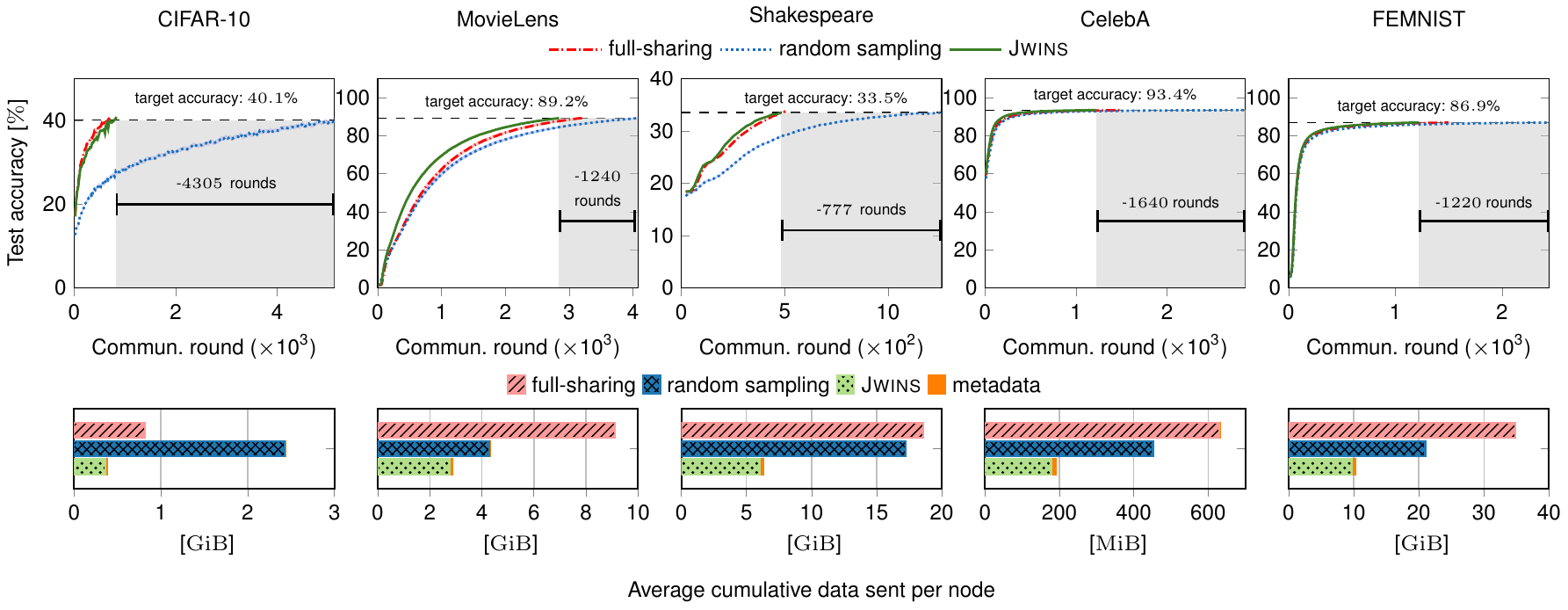}
	\caption{Learning curves and network usage for \sys compared to \fullshare and \subsampling when run until convergence.  
		In this scenario, the \subsampling algorithm is run very long and identified with a target accuracy. 
		Then \sys and \fullshare are run until this accuracy is reached. 
		\sys reaches the target accuracy much faster than \subsampling (row-1) while requiring 1.5$\times$ to 4$\times$ less network usage compared to \subsampling (row-2).}
	\label{fig:targetaccuracy_long}
\end{figure*}

\subsection{\sys Performance}
\label{sec:allComparison}
In these experiments, we run \fullshare, \subsampling, \textsc{Choco} and \sys for a fixed number of epochs.
The number of epochs for the datasets is chosen based on previous relevant work~\cite{leaf, hsiehskewscout2020, dhasade:2022:rex}.
Since we use a subsampled dataset for \shakespeare, the number of epochs is chosen to match the point where the test loss of \fullshare starts to diverge due to the models of the nodes overfitting. %
From a systems perspective, communication rounds hold more significance than epochs.
Hence, for the accuracy and loss curves, we use communication rounds on the x-axis.

\subsubsection{Learning accuracy}
\Cref{fig:sysVsDPSGDVsSubsampling} and \Cref{tab:performanceBasic} show that \sys is able to achieve as good test accuracy and test loss as \fullshare for most datasets. The performance is slightly off for the \cifar dataset, but only by 3\% in test accuracy.
However, \sys beats random sampling by 2\% to 15\% in test accuracy. 
In addition to reaching a better accuracy, \sys consistently converges faster than \subsampling.
The fact that \sys sometimes also outperforms \fullshare is due to the stochasticity of the learning process.

\subsubsection{Network savings}

Our quantifications in \Cref{tab:performanceBasic} for \Cref{fig:sysVsDPSGDVsSubsampling} show that \sys saves over 60\% of network usage across all datasets when compared to \fullshare.  
For \cifar, the difference in accuracy is  largely compensated by the \SI{397}{\gibi\byte} of network savings.
Thanks to the metadata compression, the amount of metadata in \sys is negligible compared to the model, depicted in row-3 of \Cref{fig:sysVsDPSGDVsSubsampling}.

The difference in performances across different datasets can be explained by the underlying data distributions on nodes. 
For the \celeba and \femnist datasets, the nodes likely carry samples of each class although disproportionately under the non-IID distribution.
The data partitioning in \cifar is rather extreme where nodes carry samples from only a few classes (at most 4), making it much harder to learn.
Further, the diverging test loss of \sys for \shakespeare dataset is the typical over-fitting scenario where nodes begin to over-learn on their local datasets while hurting the global test set performance.
Note that divergence is not specific to \sys and \fullshare also begins to diverge if the learning continues.

\subsubsection{\sys vs. \subsampling till convergence}
We now run another series of experiments to compare \sys and \subsampling. 
Effectively, we observe on \Cref{fig:sysVsDPSGDVsSubsampling} that \subsampling always converges slower than \fullshare and \sys.
However, the longer \subsampling is run, the more parameters are synchronized, bringing it close to \fullshare in test accuracy while also significantly increasing the total communication costs.
Therefore, when comparing \sys to \subsampling, there are two dimensions to compare against: learning performance for the same number of epochs (\Cref{sec:allComparison}), and network costs to reach the same accuracy.
We now make a fairer comparison with \subsampling in terms of network costs.
We run \subsampling for even more rounds and measure the best average accuracy reached.
We then use this accuracy as the target accuracy for \sys and \fullshare.
\Cref{fig:targetaccuracy_long} shows the number of bytes transferred for \fullshare, \subsampling, and \sys, and their convergence curves for reaching the target accuracy.
\sys always reaches this target accuracy in fewer communication rounds (row-1, \Cref{fig:targetaccuracy_long}) than \subsampling and therefore pushes between 1.5$\times$ to 4$\times$ less data on the network depending on the dataset (row-2, \Cref{fig:targetaccuracy_long}).
Interestingly, the total bytes transferred for \cifar in \subsampling is even larger than \fullshare because of the slow convergence rate of \subsampling.
Reduced communication rounds also translate to reduced wall-clock time at a similar scale.
For instance, in terms of wall-clock time for \ac{DL} training on \cifar, \sys took \SI{14}{\minute} and \subsampling took \SI{53}{\minute} to reach the same target accuracy ($3.7\times$ faster).

\subsection{\sys against \textsc{Choco}}
\label{sec:choco}
\textsc{Choco}~\cite{koloskova2019decentralized} is a state-of-the-art decentralized learning algorithm based on an error feedback mechanism that can work with multiple communication compression algorithms.
We implemented the memory-efficient \textsc{Choco}~\cite{koloskova2019decentralized} and used \topk as the compression algorithm since it worked better than \subsampling.
We challenge both \sys and \textsc{Choco} by limiting the communication budget to $20\%$ and $10\%$ of the communication budget for \fullshare.
For the communication budget of $20\%$, we specify the distribution of $\alpha$ in \sys as $p_{(\alpha = 100\%)} = 0.1$ and $p_{(\alpha = 10\%)} = 0.9$, for simplicity.
Similarly, for $10\%$ communication budget $p_{(\alpha = 100\%)} = 0.05$, $p_{(\alpha = 5\%)} = 0.95$.
\sys has the additional advantage of not introducing any new hyperparameters ($\alpha$ distribution concerns the network budget and does not require tuning). %
\textsc{Choco}, instead, introduces an additional hyperparameter: step size ($\gamma$), which needs to be tuned separately.
In our experiments, we observed that \textsc{Choco} is highly sensitive to the choice of $\gamma$.
We tuned the value of $\gamma$ to achieve the fastest convergence without crashing the learning ($\gamma_{20\%} = 0.6$ and $\gamma_{10\%} = 0.1)$.
Given the same communication budget and the exact same setting, we can also fairly compare the runtime performance of \sys against \textsc{Choco}.

\begin{figure}[t]
	\centering
	\includegraphics{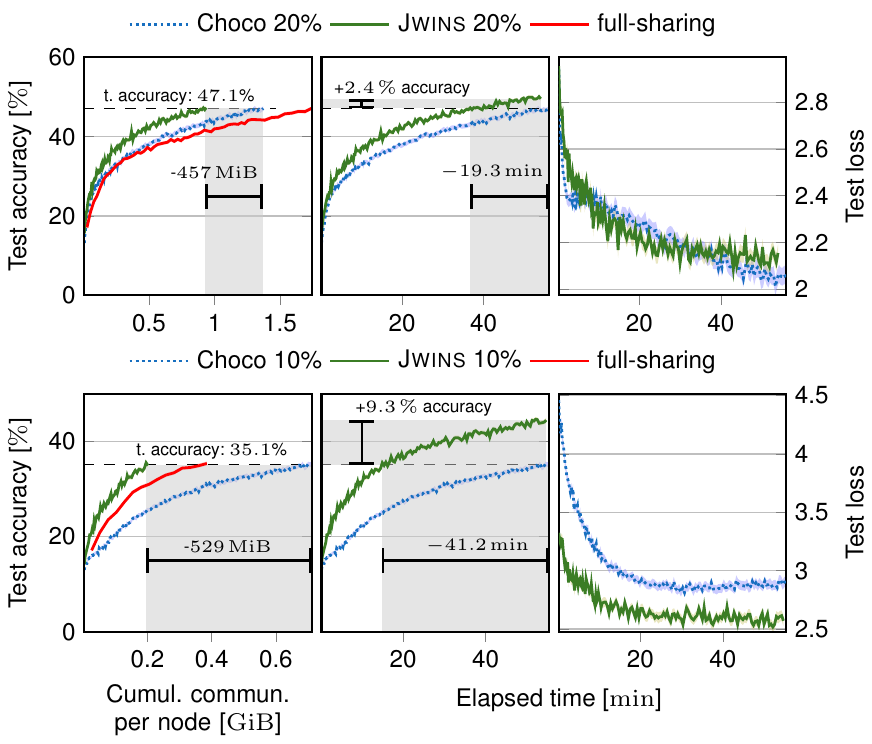}
	\caption{Performance of \sys against \textsc{Choco} for the communication budget of $20\%$ (row-1) and $10\%$ (row-2). \sys reaches the target accuracy up to $3.9\times$ faster and achieves up to $9.3\%$ better accuracy for the same communication. The performance gap gets stronger in favor of \sys as the communication budget gets smaller.}
	\label{fig:byteacc}
\end{figure}

\begin{figure}[h]
	\centering
	\includegraphics{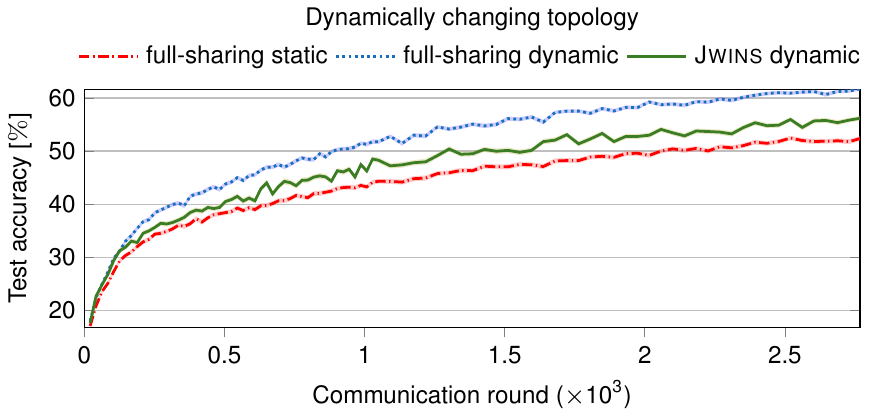}
	\caption{Randomizing neighbors every round \ie using a dynamic topology enables faster learning for both \fullshare and \sys. 
		\textsc{Choco} is unsuitable for dynamic topologies, and hence, is not charted here.}
	\label{fig:dynamic}
\end{figure}

\Cref{fig:byteacc} shows the test accuracy against the number of bytes transferred per node to reach the same accuracy, and the convergence plots when run for the same number of rounds for \sys and \textsc{Choco} over the \cifar dataset.
For completeness, we also present the communication volume of \fullshare to reach the target accuracy of \textsc{Choco}.
The total time taken by \sys to complete the same number of rounds is slightly less than \textsc{Choco}. %
This means that \sys is computationally competitive.
As \fullshare shares more data in this experiment, we cannot compare the running time of \fullshare against \textsc{Choco} and \sys in the test bed.
Even at the optimal values of $\gamma$ for the learning task, \sys converges faster than \textsc{Choco}, taking up to $3.9\times$ less time and network savings to reach the same accuracy.
\sys outperforms both \fullshare and \textsc{Choco} in terms of bytes transferred to reach a target accuracy.
As the communication budget decreases, the performance gap increases, with \sys outperforming \textsc{Choco} by $9.3\%$ accuracy.
Even with $10\%$ communication budget, \sys retains its performance against \fullshare, while \textsc{Choco} requires up to \SI{529}{\mebi\byte} less cumulative communication per node to reach the target accuracy.

Another benefit is associated with changing topologies.
Through empirical observation, we found that randomizing neighbors at each training round without moving data enables improved performance for \fullshare. 
This is due to the more effective mixing of models.
\sys demonstrates the same behavior.
Since \textsc{Choco} works on an error feedback mechanism, it is unsuitable for changing topologies and there is practically no learning in the experimental setup over a wide range of $\gamma$.
\Cref{fig:dynamic} shows the performance of \fullshare in both static and dynamic topologies, as well as \sys in dynamic topologies with the original $\alpha$ distribution on \cifar. %
Clearly, dynamic topologies result in better models for \fullshare, and \sys follows the same trend, performing even better than static-topology \fullshare.

\subsection{Breaking \sys down}
\label{sec:breakingjwins}

\begin{figure}[t]
	\centering
	\includegraphics{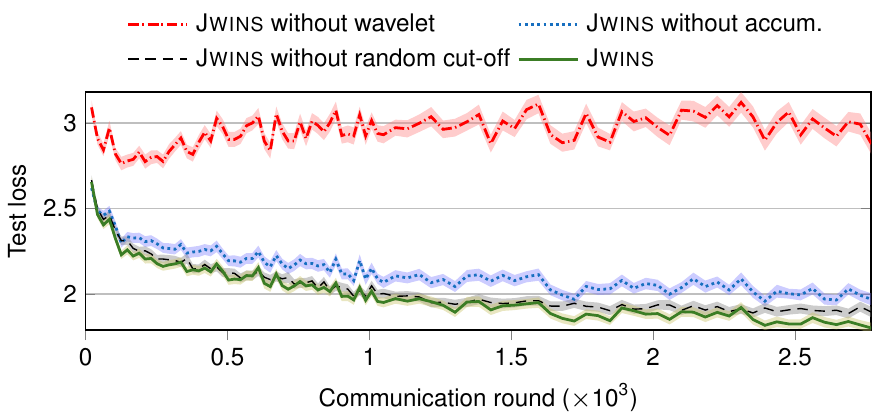}
	\caption{Ablation study of \sys on the \cifar dataset. Removing any of the three components: \emph{(i)} wavelet \emph{(ii)} accumulation or \emph{(ii)} randomized cut-off  negatively affects the performance with wavelet being the most significant.}
	\label{fig:ablationLoss}
\end{figure}

We now assess how the components of \sys impact our results in an ablation study.
We run a set of experiments by removing only one of the following components of \sys in each experiment:
(1) wavelet transform,
(2) accumulation, and
(3) randomized sharing strategy.
\Cref{fig:ablationLoss} shows the test loss of these experiments on the \cifar dataset.
It can be observed that wavelet transform is one of the most important components of \sys.
Removing wavelet transform significantly degrades the learning performance.
On the other hand, removing either accumulation or random communication cut-off causes a similar but less-significant negative impact on the performance of \sys.
When all components are put together, \sys achieves the minimum test loss.

\begin{figure}[t]
	\centering
	\includegraphics{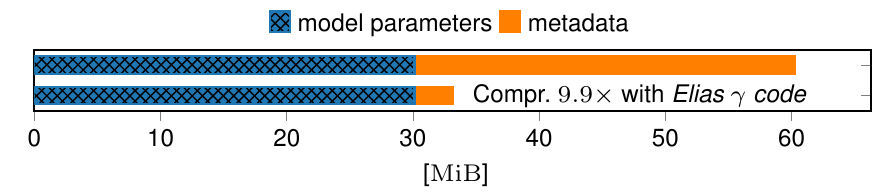}
	\caption{Size of metadata without and with compression. Without compression, approx. $50\%$ of the communication is wasted.}
	\label{fig:metadataCompression}
\end{figure}

Needless to say, metadata compression, the final component of \sys, is important in reducing network utilization.
We present the impact of metadata compression in \Cref{fig:metadataCompression} for a small 15-epoch experiment on the \cifar dataset.
The first bar is with no metadata compression, and the second bar is with Elias Gamma compression.
Without compression, the metadata is of the same size as the model parameters since both are essentially 32-bit data types.
With the Elias Gamma compression, the metadata is compressed by 9.9$\times$, allowing for more parameter sharing in the same communication budget.
This provides clear evidence that each of the components of \sys is crucial, contributing to the learning performance and network savings.

\begin{figure*}
	\centering
	\includegraphics{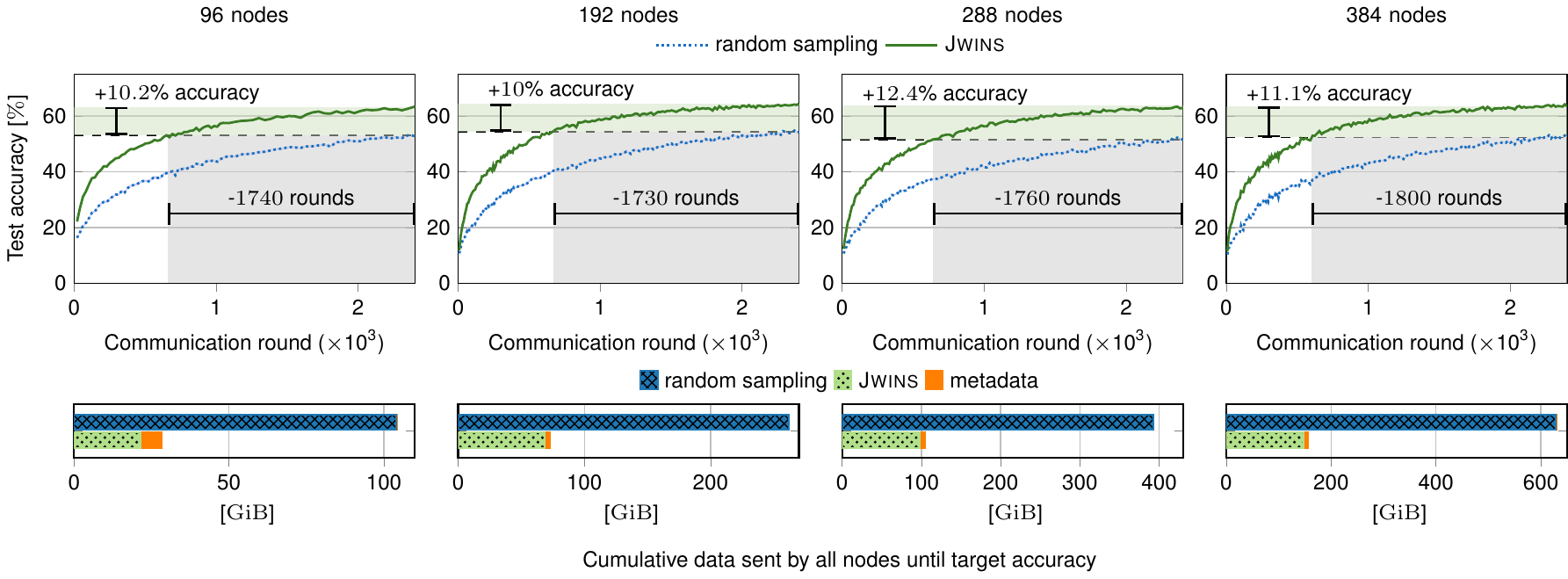}
	\caption{Scalability study for \sys on the \cifar dataset. 
		The number of nodes is increased from 96 by $2\times, 3\times$, and $4\times$.
		\sys continues to achieve higher test accuracy than \subsampling (row-1).
		In fact, with the addition of new nodes, the gross network savings of \sys amplify compared to \subsampling as more data is now exchanged (row-2, left to right). Note that the data partitioning here is less-strict \niid, hence making the values different from the ones presented in the main text.}
	\label{fig:scalabilityplus}
\end{figure*}

\subsection{Scalability of \sys}
\label{sec:scalability}

The 96-node decentralized learning is already a moderately large setting. To show that \sys can be used with a much larger number of nodes, we evaluate the scalability of \sys on the \cifar dataset with a less-strict \niid partitioning (4 shards per node instead of 2). %
We test \sys against \subsampling in 96-, 192-, 288-, and 384-node settings.
The hyperparameters for 192-, 288- and 384-nodes are tuned with the same process described in \Cref{subsec:exp_setup}.
We use a 4-regular graph for 96 nodes, 5-regular graph for 192 and 288 nodes, and 6-regular graph for 384 nodes so that the number of edges grows in proportion to the new nodes, preventing very sparsely connected graphs. 
The number of training samples per node also declines linearly with the addition of nodes since the same dataset is now partitioned across more nodes.

\Cref{fig:scalabilityplus}, row-1 depicts the evolution of test accuracy in these settings with different nodes. 
We observe that \sys consistently achieves better accuracies compared to \subsampling, also at a faster convergence rate in all scenarios.
We can also observe that \sys does not compromise on the final accuracy, even though each node has a smaller dataset as the number of nodes increases.
In fact, the gross network savings by \sys to reach a target accuracy also grow with the addition of nodes since each node now shares more data, observed in row-2 of \Cref{fig:scalabilityplus} from left to right.

\subsection{Discussion}
\label{subsec:discussion}

\paragraph{Learning performance and network savings}
In the presence of large deep learning models and also a large number of communication rounds,
\fullshare results in each node sharing up to \SI{22}{\gibi\byte} during the whole training process, as shown in \Cref{fig:sysVsDPSGDVsSubsampling}.
This makes \fullshare a highly communication-inefficient solution, despite working very well for decentralized learning tasks.

Tackling the disadvantage of \fullshare in terms of network load, \sys shares a subset of model parameters with the neighbors.
\sys saves up to 60\% in network usage, even though it adds some metadata to the messages for positional correspondence when transferring sparse vectors.

If we would allow the training to continue for an unbounded number of rounds, \fullshare would eventually achieve an accuracy slightly better than \sys.
In real-life decentralized learning tasks, this is however impractical.

\paragraph{Flexibility of \sys}
In this paper, we present \sys as a system composed of multiple modules. However, it is worth highlighting that the wavelet-based parameter ranking component of \sys holds potential for standalone use and could be integrated into other sparsification algorithms.
Furthermore, the parameter compression (Fpzip) across all experiments showed that \sys is compatible with general-purpose compression algorithms.
Finally, \sys works across a wide variety of models (CNN, LSTM, Embeddings, FC) without requiring any changes, since \sys considers models as flat vectors of parameters.

\section{Related work}
\label{sec:related}

Communication compression is a hot topic of research in distributed learning because of the large size of deep neural networks currently used.
To the best of our knowledge, \sys is the first work to use sparsification over wavelet parameter ranking and accumulation in combination with a randomized sharing strategy.

\paragraph{Communication compression in fully decentralized learning}
In \ac{DL}, quantization along with compression over the change of the model in a round has been used~\cite{tang2018communication} with convergence guarantees for unbiased compression algorithms.
On similar lines, \textsc{Choco-SGD}~\cite{koloskova2019decentralized, koloskova2020decentralized} proposes theoretical results and convergence guarantees for a wider class of compression algorithms. %
In both of these works, %
each node stores a replica of models of all other nodes to allow sharing of the compressed model-change.
Note that a memory-efficient version of \textsc{Choco-SGD} has also been proposed~\cite{koloskova2019decentralized}.
\textsc{PowerGossip} uses low-rank linear compressors applied to model difference for communication efficiency and it performs as good as \textsc{Choco-SGD} without introducing any hyperparameter.
Our work differs from these in four ways: (1) nodes in \sys do not maintain replicas of models from others, making \sys more memory-efficient, and flexible to nodes leaving and joining, (2) we propose a combination of the parameter ranking algorithm using wavelets and a random cut-off based parameter strategy (here, communication compression), (3) \sys is shown to work with a much larger number of nodes (\Cref{sec:scalability}), and (4) we evaluate \sys from a systems perspective in the presence of difficult \niid data distributions.

\vspace*{1mm}
\textsc{SAPS-PSGD}~\cite{tang2020sparsification} combines the sparsification of \subsampling with an adaptive peer-selection scheme for reducing the communication in decentralized learning with \iid data.
Partitioning schemes for logistic regression models into consecutive blocks of a given size have been proposed over \subsampling.
In contrast, our focus lies on \niid data partitioning and deep neural networks, an approach that mirrors real-world scenarios more closely and also presents a greater degree of challenge.
Moreover, \sys does not assume anything about the topology of the nodes, therefore can be combined with peer-sampling and selection services (\Cref{sec:choco}).
This is an interesting avenue for future research.

\paragraph{Compression in distributed learning}
Distributed learning with all-reduce synchronizations every round, such as the parameter server architecture, provides a much simpler platform for communicating less because every node has the same model at the start of each training round.
This allows nodes to share only gradients instead of the actual parameters.
\textsc{Gaia}~\cite{hsieh2017gaia} accumulates the local gradients in an accumulation vector and only shares them once they are higher than a certain threshold.
\textsc{DSSGD}~\cite{7447103} empirically shows that the gradients can be sparsified by up to 2 orders of magnitude in parameter-server settings using \topk.
Alistarh et al.~\cite{alistarh2018sparseconvergence} proved the convergence of \topk gradient sharing if certain assumptions are satisfied.
\topk gradient sharing is also used in \textsc{Deep Gradient Compression}~\cite{lin2018deep} along with accumulation to account for parameters that change slowly, while also using momentum correction to handle stale accumulated updates.
\textsc{SkewScout}~\cite{hsiehskewscout2020} builds on top of these communication compression techniques, like \textsc{Gaia}~\cite{hsieh2017gaia} and \textsc{Deep Gradient Compression}~\cite{lin2018deep}, by adaptively setting their hyperparameters.

All these works either rely on a central coordinator or on a global model synchronization.
In contrast, \sys is designed to work in a fully-decentralized manner, where local models of each node are not globally synchronized.

\paragraph{Wavelet transform in model compression}
Wavelet transform has been mentioned as means of data compression for ML purposes~\cite{Jin_2019, CASSIMON2020100234}.
However, to the best of our knowledge, it was never used.
\sys actually implements DWT for decentralized learning to represent the importance of parameters and wavelet proves to be quite effective. %
 
\section{Conclusion}
\label{sec:conclusion}

Overparameterized machine learning models and a large number of communication rounds in decentralized learning cause the network to be the bottleneck.
In this paper, we present \sys, a system for communication-efficient decentralized learning.
\sys maintains learning performance while reducing communication by using a smart sparsification technique, dictated by the parameter ranking and selection schemes of \sys in the wavelet-frequency domain.

Our evaluations across 5 datasets and 3 different learning tasks show that \sys can reduce the data transferred in \DL by up to $64\%$ compared to \fullshare without deteriorating the performance.
Further, \sys achieves the convergence accuracy of its \subsampling competitor significantly faster while transferring up to $4\times$ fewer bytes.
We also show that the parameter ranking in \sys is highly effective through an ablation study.

Interesting avenues of future research can be along both dimensions of parameter ranking and parameter selection.
An adaptive version of the importance score based on the parameter type (CNN, RNN, FC) may be explored in depth.
Theoretical convergence guarantees of \sys would complement our empirical evaluations.
Parameter selection based on inferred knowledge about the models of other participating nodes is another future research direction.

\printbibliography

@misc{boccassi:2022:zmq,
	author={Luca Boccassi and others},
	title={ZeroMQ: An open-source universal messaging library},
	url={https://zeromq.org},
	year={2022},
}

@inproceedings{li:2014:parameterserver,
	title	= {Scaling Distributed Machine Learning with the Parameter Server},
	author	= {Mu Li and David G. Anderson and Jun Woo Park and Alexander J. Smola and Amr Ahmed and Vanja Josifovski and James Long and Eugene J. Shekita and Bor-Yiing Su},
	year	= {2014},
	booktitle	= {OSDI},
    url={https://www.usenix.org/system/files/conference/osdi14/osdi14-paper-li_mu.pdf},
}

@misc{collet:lz4:2022,
	author={Yann Collet},
	title={{LZ4}},
	url={http://www.lz4.org/},
	year={2022},
}

@misc{pavlov:lzma:2022,
	author={Igor Pavlov},
	title={{LZMA} {SDK}},
	url={https://www.7-zip.org/},
	year={2022},
}

@article{ormandi:2013:gossiplearning,
	author = {Ormándi, Róbert and Hegedűs, István and Jelasity, Márk},
	title = {Gossip learning with linear models on fully distributed data},
	journal = {Concurrency and Computation: Practice and Experience},
	volume = {25},
	number = {4},
	doi = {10.1002/cpe.2858},
	year = {2013}
}

@misc{he2015delving,
      title={Delving Deep into Rectifiers: Surpassing Human-Level Performance on ImageNet Classification}, 
      author={Kaiming He and Xiangyu Zhang and Shaoqing Ren and Jian Sun},
      year={2015},
      eprint={1502.01852},
      archivePrefix={arXiv},
}

@misc{hannun2014deep,
      title={Deep Speech: Scaling up end-to-end speech recognition}, 
      author={Awni Hannun and Carl Case and Jared Casper and Bryan Catanzaro and Greg Diamos and Erich Elsen and Ryan Prenger and Sanjeev Satheesh and Shubho Sengupta and Adam Coates and Andrew Y. Ng},
      year={2014},
      eprint={1412.5567},
      archivePrefix={arXiv},
}

@INPROCEEDINGS{8237349,
      author={He, Wenhao and Zhang, Xu-Yao and Yin, Fei and Liu, Cheng-Lin},
      booktitle={2017 IEEE International Conference on Computer Vision (ICCV)}, 
      title={Deep Direct Regression for Multi-oriented Scene Text Detection}, 
      year={2017},
      volume={},
      number={},
      pages={745-753},
      doi={10.1109/ICCV.2017.87}
}

@inproceedings{mcmahan2017communicationefficient,
    title = 	 {Communication-Efficient Learning of Deep Networks from Decentralized Data},
    author = 	 {McMahan, Brendan and Moore, Eider and Ramage, Daniel and Hampson, Seth and Arcas, Blaise Aguera y},
    year = 	 {2017},
    booktitle = 	 {AISTATS},
    url = 	 {https://proceedings.mlr.press/v54/mcmahan17a.html},
}

@inproceedings{lian2017dpsgd,
	author = {Lian, Xiangru and Zhang, Ce and Zhang, Huan and Hsieh, Cho-Jui and Zhang, Wei and Liu, Ji},
	title = {Can Decentralized Algorithms Outperform Centralized Algorithms? A Case Study for Decentralized Parallel Stochastic Gradient Descent},
	year = {2017},
	booktitle = {NIPS},
	url = {https://proceedings.neurips.cc/paper/2017/file/f75526659f31040afeb61cb7133e4e6d-Paper.pdf},
}

@INPROCEEDINGS{7447103,
    author={Shokri, Reza and Shmatikov, Vitaly},
    booktitle={2015 53rd Annual Allerton Conference on Communication, Control, and Computing (Allerton)}, 
    title={Privacy-preserving deep learning}, 
    year={2015},
    volume={},
    number={},
    pages={909-910},
    doi={10.1109/ALLERTON.2015.7447103}
}

@article{HEGEDUS2021109,
    title = {Decentralized learning works: An empirical comparison of gossip learning and federated learning},
    journal = {Journal of Parallel and Distributed Computing},
    volume = {148},
    pages = {109-124},
    year = {2021},
    issn = {0743-7315},
    doi = {10.1016/j.jpdc.2020.10.006},
    author = {István Hegedűs and Gábor Danner and Márk Jelasity}
}

@misc{leaf,
      title={LEAF: A Benchmark for Federated Settings}, 
      author={Sebastian Caldas and Sai Meher Karthik Duddu and Peter Wu and Tian Li and Jakub Konečný and H. Brendan McMahan and Virginia Smith and Ameet Talwalkar},
      year={2019},
      eprint={1812.01097},
      archivePrefix={arXiv},
}

@inproceedings{hsiehskewscout2020,
    author = {Hsieh, Kevin and Phanishayee, Amar and Mutlu, Onur and Gibbons, Phillip B.},
    title = {The Non-{IID} Data Quagmire of Decentralized Machine Learning},
    year = {2020},
    url={http://proceedings.mlr.press/v119/hsieh20a/hsieh20a.pdf},
    booktitle = {ICML}
}

@article{korenmatrixfactorization2009,
    author = {Koren, Yehuda and Bell, Robert and Volinsky, Chris},
    title = {Matrix Factorization Techniques for Recommender Systems},
    year = {2009},
    issue_date = {August 2009},
    address = {Washington, DC, USA},
    volume = {42},
    number = {8},
    issn = {0018-9162},
    doi = {10.1109/MC.2009.263},
    journal = {Computer},
    month = {aug},
    pages = {30--37},
    numpages = {8},
}

@article{krizhevsky2014cifar,
    title={The CIFAR-10 dataset},
    author={Krizhevsky, Alex and Nair, Vinod and Hinton, Geoffrey},
    url={https://www.cs.toronto.edu/~kriz/cifar.html},
    volume={55},
    number={5},
    year={2014}
}

@article{movielensdataset2015,
    author = {Harper, F. Maxwell and Konstan, Joseph A.},
    title = {The MovieLens Datasets: History and Context},
    year = {2015},
    issue_date = {January 2016},
    address = {New York, NY, USA},
    volume = {5},
    number = {4},
    issn = {2160-6455},
    doi = {10.1145/2827872},
    journal = {ACM Trans. Interact. Intell. Syst.},
    month = {dec},
    articleno = {19},
    numpages = {19},
}

@inproceedings{tang2020sparsification,
  author={Tang, Zhenheng and Shi, Shaohuai and Chu, Xiaowen},
  booktitle={2020 IEEE 40th International Conference on Distributed Computing Systems (ICDCS)}, 
  title={Communication-Efficient Decentralized Learning with Sparsification and Adaptive Peer Selection}, 
  year={2020},
  doi={10.1109/ICDCS47774.2020.00153}}

@inproceedings{koloskova2019decentralized,
  title={Decentralized stochastic optimization and gossip algorithms with compressed communication},
  author={Koloskova, Anastasia and Stich, Sebastian and Jaggi, Martin},
  year={2019},
  url={https://proceedings.mlr.press/v97/koloskova19a.html},
  booktitle={ICML},
}

@inproceedings{koloskova2020decentralized,
  title={Decentralized Deep Learning with Arbitrary Communication Compression},
  author={Koloskova, Anastasia and Lin, Tao and Stich, Sebastian U and Jaggi, Martin},
  booktitle={ICLR},
  url={https://openreview.net/pdf?id=SkgGCkrKvH},
  year={2020}
}

@InProceedings{pmlr-v119-koloskova20a,
    title = 	 {A Unified Theory of Decentralized {SGD} with Changing Topology and Local Updates},
    author =       {Koloskova, Anastasia and Loizou, Nicolas and Boreiri, Sadra and Jaggi, Martin and Stich, Sebastian},
    booktitle = 	 {ICML},
    year = 	 {2020},
    url = 	 {https://proceedings.mlr.press/v119/koloskova20a.html},
}

@inproceedings{dhasade:2023:dcpy,
    author={Dhasade, Akash 
        and Kermarrec, Anne-Marie 
        and Pires, Rafael 
        and Sharma, Rishi 
        and Vujasinovic, Milos},
    booktitle={3rd Workshop on Machine Learning and Systems},
    title={Decentralized Learning Made Easy with {DecentralizePy}},
    doi={10.1145/3578356.3592587},
    year={2023},
}

@inproceedings{dhasade:2022:rex,
	author={Dhasade, Akash
	and Dresevic, Nevena
	and Kermarrec, Anne-Marie
	and Pires, Rafael},
	booktitle={36th {IEEE} International Parallel and Distributed Processing Symposium ({IPDPS})}, 
	title={{TEE}-based decentralized recommender systems: The raw data sharing redemption},
	doi={10.1109/IPDPS53621.2022.00050},
	year={2022},
}

@incollection{torchSoftware,
    title = {PyTorch: An Imperative Style, High-Performance Deep Learning Library},
    author = {Paszke, Adam and Gross, Sam and Massa, Francisco and Lerer, Adam and Bradbury, James and Chanan, Gregory and Killeen, Trevor and Lin, Zeming and Gimelshein, Natalia and Antiga, Luca and Desmaison, Alban and Kopf, Andreas and Yang, Edward and DeVito, Zachary and Raison, Martin and Tejani, Alykhan and Chilamkurthy, Sasank and Steiner, Benoit and Fang, Lu and Bai, Junjie and Chintala, Soumith},
    booktitle = {NeurIPS},
    year = {2019},
    url = {http://papers.neurips.cc/paper/9015-pytorch-an-imperative-style-high-performance-deep-learning-library.pdf}
}

@article{lindstrom2006fpzip,
    author = {Lindstrom, Peter and Isenburg, Martin},
    title = {Fast and Efficient Compression of Floating-Point Data},
    year = {2006},
    issue_date = {September 2006},
    address = {USA},
    volume = {12},
    number = {5},
    issn = {1077-2626},
    doi = {10.1109/TVCG.2006.143},
    journal = {IEEE Transactions on Visualization and Computer Graphics},
    month = {sep},
    pages = {1245--1250},
    numpages = {6},
}

@article{XIAO200465,
	title = {Fast linear iterations for distributed averaging},
	journal = {Systems \& Control Letters},
	volume = {53},
	number = {1},
	pages = {65-78},
	year = {2004},
	issn = {0167-6911},
	doi = {10.1016/j.sysconle.2004.02.022},
	author = {Lin Xiao and Stephen Boyd},
}

@article{elias1975universal,
  title={Universal codeword sets and representations of the integers},
  author={Elias, Peter},
  journal={IEEE transactions on information theory},
  volume={21},
  number={2},
  pages={194--203},
  year={1975},
  doi={10.1109/TIT.1975.1055349},
  publisher={IEEE}
}

@software{Lee2019,
    author       = {Gregory Lee and
            Ralf Gommers and
            Kai Wohlfahrt and
            Filip Wasilewski and
            Aaron O'Leary and
            Holger Nahrstaedt and
            Alexandre Sauvé and
            Ankit Agrawal and
            Daniel M. Pelt and
            Helder Oliveira and
            Thomas Arildsen and
            Frank Yu and
            Matthew Brett and
            Michel Pelletier and
            SylvainLan and
            Daniele Tricoli and
            Saket Choudhary and
            asnt and
            Arfon Smith and
            0-tree and
            Balint Reczey and
            Corey Goldberg and
            Daniel Goertzen and
            Dawid Laszuk and
            ElConno and
            Jacopo Antonello and
            Jakub Mandula and
            jakirkham and
            Jonathan Dan and
            Lokesh Ravindranathan},
    title        = {PyWavelets/pywt: PyWavelets 1.3.0},
    month        = mar,
    year         = 2022,
    version      = {v1.3.0},
    doi          = {10.5281/zenodo.6347505},
}

@inproceedings{seide2014-bit,
    author = {Seide, Frank and Fu, Hao and Droppo, Jasha and Li, Gang and Yu, Dong},
    title = {1-Bit Stochastic Gradient Descent and Application to Data-Parallel Distributed Training of Speech {DNNs}},
    booktitle = {Interspeech 2014},
    year = {2014},
    month = {September},
    url={https://www.microsoft.com/en-us/research/wp-content/uploads/2016/02/IS140694.pdf},
}

@inproceedings{alistarh2017quantization,
    author = {Alistarh, Dan and Grubic, Demjan and Li, Jerry Z. and Tomioka, Ryota and Vojnovic, Milan},
    title = {{QSGD}: Communication-Efficient {SGD} via Gradient Quantization and Encoding},
    year = {2017},
    url={https://proceedings.neurips.cc/paper_files/paper/2017/file/6c340f25839e6acdc73414517203f5f0-Paper.pdf},
    isbn = {9781510860964},
    booktitle = {NIPS}
}

@inproceedings{ lin2018deep,
    title={Deep Gradient Compression: Reducing the Communication Bandwidth for Distributed Training},
    author={Yujun Lin and Song Han and Huizi Mao and Yu Wang and Bill Dally},
    booktitle={ICLR},
    year={2018},
    url={https://openreview.net/forum?id=SkhQHMW0W},
}

@inproceedings{strom2015scalable,
    title={Scalable distributed {DNN} training using commodity {GPU} cloud computing},
    author={Strom, Nikko},
    booktitle={16th Annual Conference of the International Speech Communication Association},
    url={https://www.isca-speech.org/archive_v0/interspeech_2015/papers/i15_1488.pdf},
    year={2015}
}

@inproceedings{aji-heafield-2017-sparse,
    title = "Sparse Communication for Distributed Gradient Descent",
    author = "Aji, Alham Fikri  and
      Heafield, Kenneth",
    booktitle = "2017 Conference on Empirical Methods in Natural Language Processing",
    year = "2017",
    doi = "10.18653/v1/D17-1045",
}

@inproceedings {hsieh2017gaia,
    author = {Kevin Hsieh and Aaron Harlap and Nandita Vijaykumar and Dimitris Konomis and Gregory R. Ganger and Phillip B. Gibbons and Onur Mutlu},
    title = {Gaia: {Geo-Distributed} Machine Learning Approaching {LAN} Speeds},
    booktitle = {NSDI},
    year = {2017},
    isbn = {978-1-931971-37-9},
    url = {https://www.usenix.org/conference/nsdi17/technical-sessions/presentation/hsieh},
}

@inproceedings{alistarh2018sparseconvergence,
    author = {Alistarh, Dan and Hoefler, Torsten and Johansson, Mikael and Khirirat, Sarit and Konstantinov, Nikola and Renggli, C\'{e}dric},
    title = {The Convergence of Sparsified Gradient Methods},
    url={https://proceedings.neurips.cc/paper_files/paper/2018/file/314450613369e0ee72d0da7f6fee773c-Paper.pdf},
    year = {2018},
    location = {Montr\'{e}al, Canada},
    booktitle = {NeurIPS}
}

@inproceedings{tang2018communication,
  title={Communication compression for decentralized training},
  author={Tang, Hanlin and Gan, Shaoduo and Zhang, Ce and Zhang, Tong and Liu, Ji},
  booktitle={NeurIPS},
  url={https://papers.nips.cc/paper_files/paper/2018/file/44feb0096faa8326192570788b38c1d1-Paper.pdf},
  year={2018}
}

@article{CASSIMON2020100234,
    title = {Designing resource-constrained neural networks using neural architecture search targeting embedded devices},
    journal = {Internet of Things},
    volume = {12},
    pages = {100234},
    year = {2020},
    issn = {2542-6605},
    doi = {10.1016/j.iot.2020.100234},
    author = {Thomas Cassimon and Simon Vanneste and Stig Bosmans and Siegfried Mercelis and Peter Hellinckx},
}

@inproceedings{Jin_2019,
	doi = {10.1145/3307681.3326608},
	year = 2019,
	month = {jun},
	author = {Sian Jin and Sheng Di and Xin Liang and Jiannan Tian and Dingwen Tao and Franck Cappello},
	title = {{DeepSZ}: A Novel Framework to Compress Deep Neural Networks by Using Error-Bounded Lossy Compression},
	booktitle = {28th International Symposium on High-Performance Parallel and Distributed Computing}
}

@inproceedings{mantoro2017comparison,
  title={Comparison methods of {DCT}, {DWT} and {FFT} techniques approach on lossy image compression},
  author={Mantoro, Teddy and Alfiah, Fifit},
  booktitle={2017 International Conference on Computing, Engineering, and Design (ICCED)},
  doi={10.1109/CED.2017.8308126},
  pages={1--4},
  year={2017},
}

@article{ramkumar1997fft,
    title={An {FFT}-based technique for fast fractal image compression},
    author={Ramkumar, Mahalingam and Anand, GV},
    journal={Signal processing},
    volume={63},
    number={3},
    year={1997},
    doi={10.1016/S0165-1684(97)00162-X},
    publisher={Elsevier}
}

@inproceedings{patel2016improved,
    title={An improved image compression technique using Huffman coding and {FFT}},
    author={Patel, Rachit and Katiyar, Sapna and Arora, Khushboo},
    booktitle={International conference on smart trends for information technology and computer communications},
    year={2016},
    doi={10.1007/978-981-10-3433-6_7},
    organization={Springer}
}

@article{taubman2002jpeg2000,
    title={{JPEG2000}: Standard for interactive imaging},
    author={Taubman, David S and Marcellin, Michael W},
    journal={Proceedings of the IEEE},
    volume={90},
    number={8},
    pages={1336--1357},
    year={2002},
    doi={10.1109/JPROC.2002.800725},
    publisher={IEEE}
}

@inproceedings{vogels2020practical,
  title={Practical low-rank communication compression in decentralized deep learning},
  author={Vogels, Thijs and Karimireddy, Sai Praneeth and Jaggi, Martin},
  url={https://proceedings.neurips.cc/paper_files/paper/2020/file/a376802c0811f1b9088828288eb0d3f0-Paper.pdf},
  booktitle={NeurIPS},
  year={2020}
}

\end{document}